\documentclass[preprint,times,12pt]{elsarticle}

\usepackage{graphicx}
\usepackage{dcolumn}
\usepackage{multirow}
\usepackage{amsmath,bm}
\usepackage{color}%
\usepackage{pdflscape}

\begin{document}

\title{Strain Effects on Phase Transitions in Transition Metal
Dichalcogenides}

\author{Seoung-Hun Kang}
\author{Young-Kyun Kwon\corref{cor}}
\ead{ykkwon@khu.ac.kr}
\cortext[cor]{Corresponding author.}
\address{Department of Physics and
             Research Institute for Basic Sciences,
             Kyung Hee University, Seoul, 02447, Korea}
\address{Korea Institute for Advanced Study (KIAS), Seoul 02455, Korea}

\date{\today}

\begin{abstract}
We perform density functional theory calculation to investigate the
structural and electronic properties of various two-dimensional
transition metal dichalcogenides, MX$_2$ (M$=$Ti, V, Cr, Zr, Nb, Mo,
Hf, Ta, or W, and X$=$S or Se), and their strain-induced phase
transitions. We evaluate the relative stability and the activation
barrier between the octahedral-\textit{T} and the trigonal-\textit{H}
phases of each MX$_2$. It is found that the equilibrium and phase
transition characteristics of MX$_2$ can be classified by the group to
which its metal element M belongs in the periodic table. MX$_2$ with M
in the group 4 (Ti, Zr, or Hf), forms an octahedral-\textit{T} phase,
while that with an M in the group 6 (Cr, Mo, or W) does a
trigonal-\textit{H} phase. On the other hand, MX$_2$ with M in the
group 5 (V, Nb, Ta), which is in-between the groups 4 and 6, may form
either phase with a similar stability. It is also found that their
electronic structures are strongly correlated to the structural
configurations: mostly metallic in the \textit{T} phase, while
semiconducting in the \textit{H} phase, although there are some
exceptions. We also explore the effects of an applied stress and find
for some MX$_2$ materials that the resultant strain, either tensile or
compressive, may induce a structural phase transition by reducing the
transition energy barrier, which is, in some cases, accompanied by its
metal-insulator transition.
\end{abstract}
\begin{keyword}
Transition metal dichalcogenides; Phase transition; Strain effect;
Density functional theory
\end{keyword}


\maketitle

\section{Introduction}
\label{Introduction}

Transition metal dichalcogenides (TMDCs) in the form of a monolayer
have emerged as they have been easily exfoliated to a single layer
due to the weak interlayer interaction like in graphene. One of the
unique properties of TMDCs is polymorphism with distinct electronic
properties. The relative arrangement of chalcogen atoms to the
transition metal atom determines the structural symmetry, and thus
the phase, trigonal prismatic \textit{H} (D$_{3h}$) or octahedral
\textit{T} phase (O$_h$). Although there are a variety of combinations
in TMDCs, most TMDC studies have focused mainly on MoS$_2$. It was
interestingly revealed that MoS$_2$ can have completely different
electronic characteristics that are semiconducting and metallic in the
\textit{H} and \textit{T} phases, respectively.~\cite{{Mattheiss1973},
{Schollhorn1992},{Kannewurf1993}} It was also observed in MoS$_2$
that both \textit{H} and \textit{T} phases could coexist in a certain
condition,~\cite{Eda2012} and metal-insulator transitions were induced
by strain.~\cite{{Guo2015},{Kan2014}} Furthermore, it was reported
that MoS$_2$ has many useful properties such as electronic, optical,
mechanical, and chemical properties as well as thermal
properties,~\cite{{Yoffe1969},{Yoffe1973},{Yoffe1993}} which were
demonstrated by MoS$_2$-based electrical and photoelectronic
devices.~\cite{{Kis2011},{Strano2012}} Some other TMDC materials, such
as MoSe$_2$, WS$_2$, WSe$_2$, were also used to study their various
physical properties.~\cite{{Strano2012},{Zhang2013}}

Recent experimental and theoretical studies have endeavored to control
the electronic property of MoS$_2$ through an electron doping from
intercalated alkali atoms,~\cite{{Schollhorn1992}, {Chhowalla2011},
{Kanatzidis1999},{Weber1999},{Ouyang2015}}, in-plane
strain~\cite{{Kan2014},{Ouyang2015}} or the substitutional doping of
Re atoms.~\cite{Suenaga2014} Although it has been reported that the
MoS$_2$ band gap can be changed by strain~\cite{Stesmans2011} or
hydrostatic pressure,~\cite{Lin2014} there have been no systematic
studies on the electronic characteristics of various single-layered
TMDCs affected by structural deformations such as strain.

In this paper, we report a first-principles study of the structural
and electronic properties of MX$_2$ (M$=$Ti, V, Cr, Zr, Nb, Mo, Hf,
Ta, or W, and X$=$S or Se) and their phase transitions. 
Our study revealed that each MX$_2$ has a preferred phase
or either phase depending on the group to which its metal atom M
belongs in the periodic table. For the phase transitions, we evaluated
the relative stability during phase transition between two different
phases (octahedral-\textit{T} and trigonal-\textit{H}) of each
composition, mimicked by the intralayer gliding of one of the
chalcogen atom X planes relative to the metal atom M plane. We also
explored the external strain effects on the phase transition by
considering changes in activation energy barriers and reaction
energies. We found that the external tensile strain may induce the
phase transition in MX$_2$ with M in the group 6.

\section{Computational details}
\label{Computational}

To systematically investigate the structural stabilities and the
electronic properties of various single-layered TMDCs, and the strain
effect on their phase transitions, we used \emph{ab initio} density
functional theory (DFT)~\cite{Kohn1965} as implemented in the Vienna
\emph{ab initio} simulation packages (VASP).~\cite{{Kresse1996},
{Kresse1993}} For the exchange-correlation functional, we used
the generalized gradient approximation (GGA) with the
Perdew-Burke-Ernzerhof parameterization~\cite{PBE} with spin
polarization. 
The local density approximation (LDA) was also used for comparison.
Valence electrons were described under the projector
augmented wave potentials,~\cite{PAW} and the electronic wave
functions were expanded by a plane wave basis set with a cutoff energy
of 400~eV. The corresponding Brillouin zone was sampled using a
$\Gamma$-centered $50{\times}50{\times}1$ $k$-grid mesh for the
primitive unit cell. The charge density was determined
self-consistently with a precision of ${\le}10^{-5}$~eV/cell in total
energy. A vacuum region of ${\sim}22$~{\AA} was included in a unit
cell along a direction perpendicular to the MX$_2$ plane to avoid the
slab-slab interaction with the neighboring cells. All geometrical
relaxations were continued until the Helmann-Feynman force acting on
every atom had become lower than 0.01~eV/{\AA}. For heavy elements,
such as Hf, Ta, and W, we took into account the spin-orbit coupling (SOC) to describe their accurate electronic properties. 

\section{Results and discussion}
\label{Results}

\begin{figure}[t]
\includegraphics[width=1.0\columnwidth]{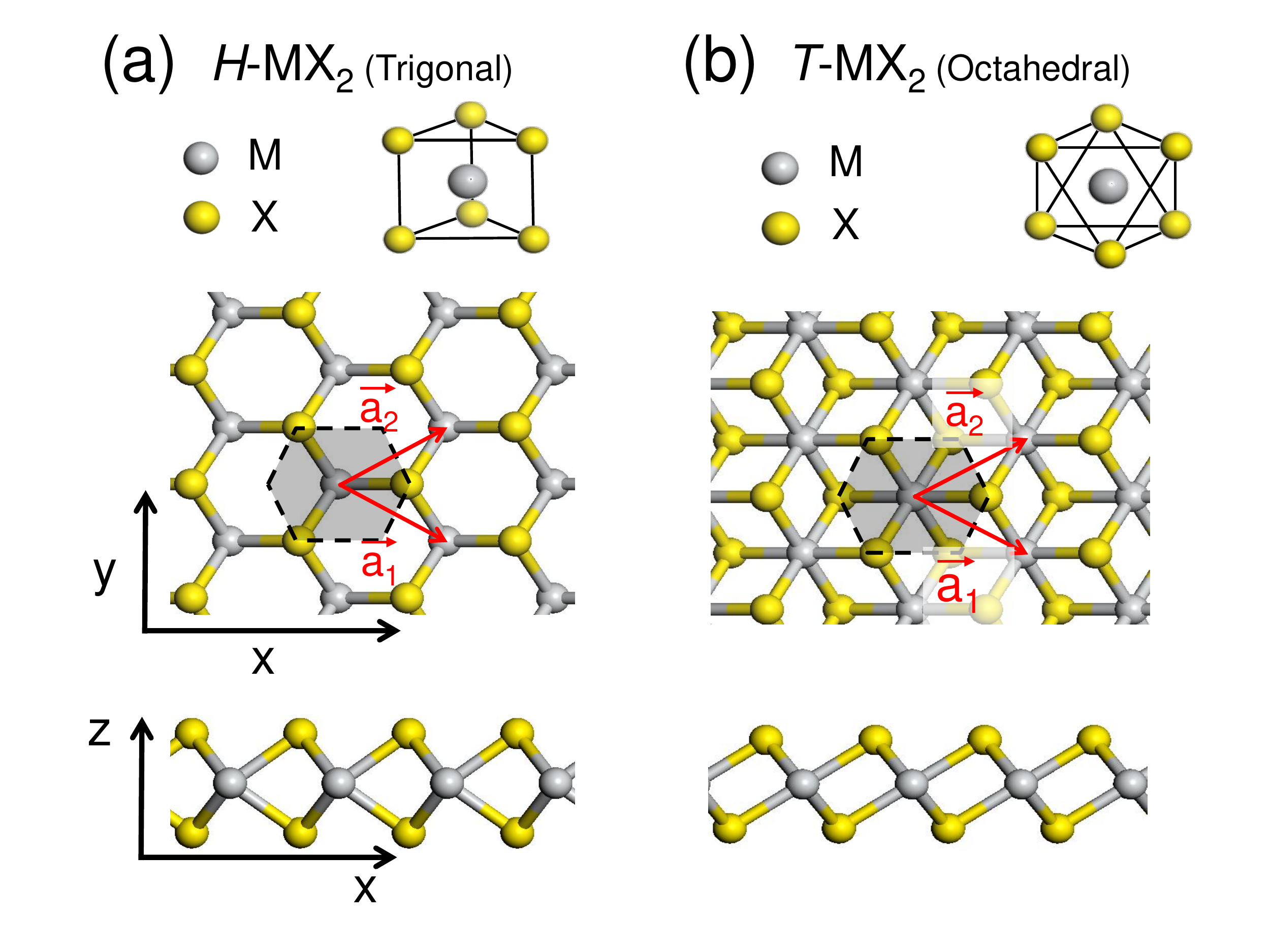}
\caption{(Color online) 
Two different structural configurations of MX$_2$: (a) trigonal
prismatic (\textit{H} phase) and (b) octahedral (\textit{T} phase)
configurations in basal plane and cross-section views together with
their corresponding symmetries. Both \textit{H} and \textit{T}
configurations form a hexagonal lattice composed of three atom basis
of one transition metal atom M in the middle and two chalcogen atoms X
and X$^\prime$ on the sides with ``ABA'' and ``ABC'' atomic stacking 
sequences (X-M-X$^\prime$) in the Wigner-Seitz cell denoted by a gray-
shaded region enclosed by a dashed hexagon. Two dimensional lattice 
vectors $\mathbf{a}_1$ and $\mathbf{a}_2$ are indicated by red
arrows. Gray and yellow spheres depict transition metal (M) atoms and
chalcogen (X) atoms, respectively.
\label{Fig1}}
\end{figure}

It has been known that the TMDCs have a characteristic layered
structure and can form different structural types due to the complex
registry of atomic stacking sequence of metal and chalcogen
atoms.~\cite{Kertesz1984} To perform a comparative study, we
considered 18 different single-layered TMDCs. Each TMDC or MX$_2$
layer is composed of a plane of a transition metal element M (M$=$Ti,
V, Cr, Zr, Nb, Mo, Hf, Ta, or W) sandwiched by two planes of chalcogen
element X (X$=$S or Se). Depending on how to stack two chalcogen
planes relative to each other, there are two representative possible
configurations, trigonal-prismatic (\textit{H}) and octahedral
(\textit{T}) phases, which are known to be mostly observed in MX$_2$
materials,~\cite{Kertesz1984} as shown in Fig.~\ref{Fig1}. The
\textit{H} and \textit{T} phases are in resemblance to ``ABA'' and
``ABC'' stacking sequences in a stacked triangular lattice.
Figure~\ref{Fig1} clearly shows a difference between two phases. In
the basal plane view of Fig.~\ref{Fig1}, both phases form a honeycomb
lattice, which contains one M and two X atoms in the Wigner-Seitz
cell.

\begin{figure*}[t]
\includegraphics[width=1.0\textwidth]{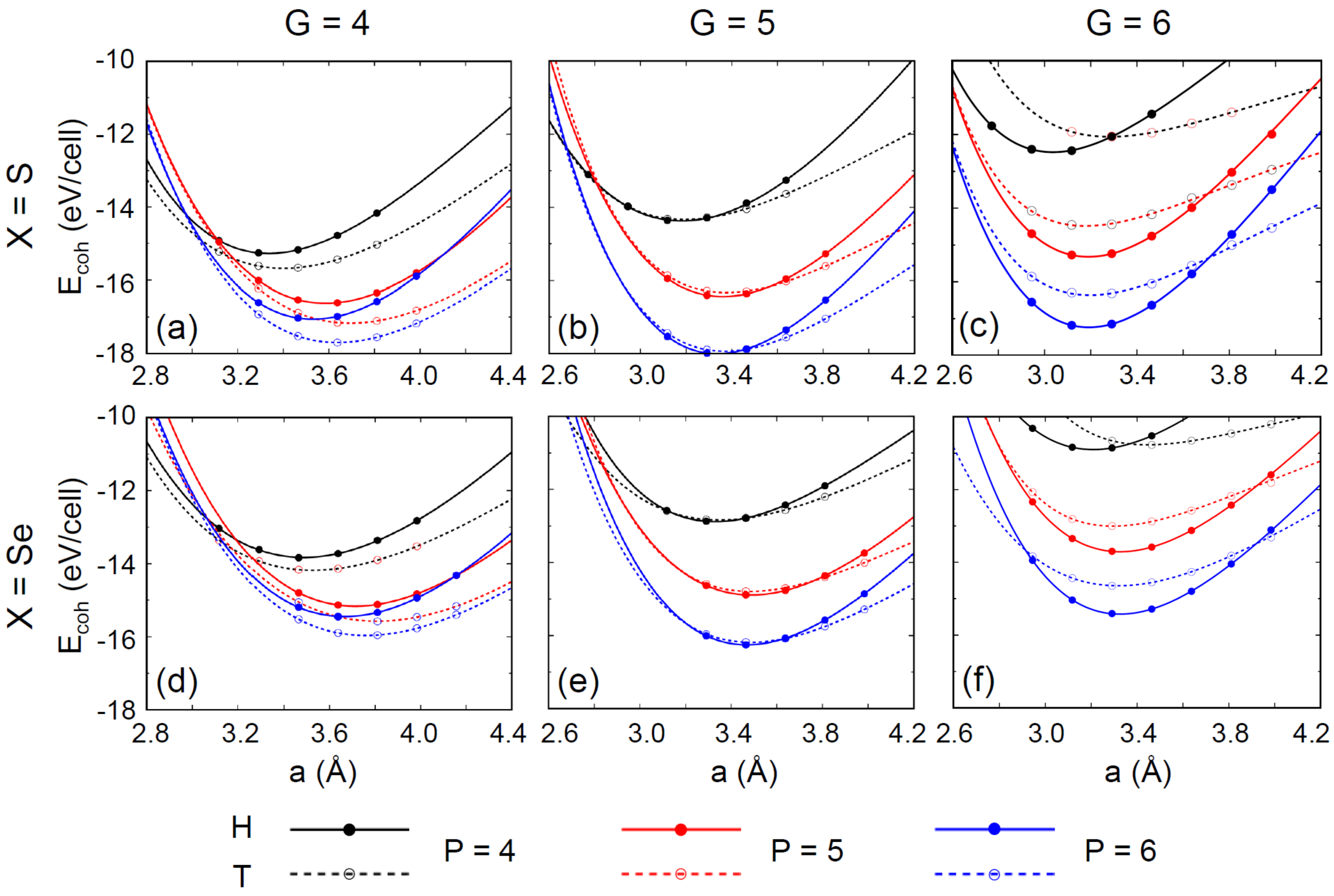}
\caption{(Color online) Cohesive energy $E_{\textrm{coh}}$ of
M$_G^P$X$_2$ as a function of lattice constants. Here, the
superscript $P$ and the subscript $G$ denote respectively the period
and the group to which M belongs in the periodic table: (a)
M$_4^P$S$_2$ (b) M$_5^P$S$_2$, (c) M$_6^P$S$_2$, (d) M$_4^P$Se$_2$,
(e) M$_5^P$Se$_2$, and (f) M$_6^P$Se$_2$. In each figure, black, red,
and blue lines represent $P=4$, 5, and 6, corresponding to
M$_4^P=$(Ti, Zr, Hf), M$_5^P=$(V, Nb, Ta), and M$_6^P=$(Cr, Mo, W),
respectively; and solid and dotted lines delineate $E_{\textrm{coh}}$
of the \textit{H} and \textit{T} phases for a given MX$_2$, as
indicated by legend in the bottom.
\label{Fig2}}
\end{figure*}

For clear description of 18 different TMDC materials and their
systematic analysis, we denoted each TMDC by a notation of
M$_G^P$X$_2$, where the superscript $P$ and the subscript $G$ denote
the period and the group to which the transition metal element M
belongs in the periodic table, and X is either sulfur (S) or selenium
(Se) categorizing into sulfide or selenide. We further grouped the
respective chalcogenides in terms of groups $G$ of M: M$_4^P=$Ti, Zr,
Hf for $P=4, 5, 6$, respectively, and similarly M$_5^P=$V, Nb, Ta, and
M$_6^P=$Cr, Mo, W. Figure~\ref{Fig2} shows the calculated cohesive
energies of the \textit{H} and \textit{T} phases of M$_G^P$X$_2$ as a
function of lattice constants: (a), (b), and (c) are for sulfides,
while (d), (e), and (f) are for selenides, of M$_4^P$, M$_5^P$, and
M$_6^P$, respectively, for three different $P$ values. Each energy
point in Fig.~\ref{Fig2} was obtained by finding the equilibrium
structure for the corresponding lattice constant.

We found an overall common trend from Fig.~\ref{Fig2} that TMDCs
with a heavier TM element appear to be more stable than those with a
lighter one within the same group $G$. It was, moreover, found that
the group $G$, to which the transition metal atom M belongs,
determines the equilibrium phases of M$_G^P$X$_2$ materials. As
clearly shown in Fig.~\ref{Fig2} (a) and (d), M$_4^P$X$_2$ materials
prefer the octahedral \textit{T} phase (dotted line) to the \textit{H}
counterpart (solid line). On the other hand, the materials with a
metal atom M in the group 6 (Cr, Mo, and W) favor the trigonal
\textit{H} phase~\cite{Yoffe1969} over the \textit{T} counterpart, as
shown in Fig.~\ref{Fig2} (c) and (f). Intriguingly, the TMDCs with the
group 5 metal atom (V, Nb, and Ta) may coexist in both
phases~\cite{Yoffe1969} as displayed in Fig.~\ref{Fig2} (b) and (e).
This result is consistent with a previous work finding coexistence of VS$_2$ in both phases.~\cite{Kan2014} These 
results indicate that the stable phase of each TMDC is mainly
determined by the group to which the transition metal belongs in the
periodic table, as verified by the electronic occupation in its $d$
orbitals.~\cite{Zhang2013} 
It is worth noting that the structural stability can be determined by free energy rather than cohesive energy. Nevertheless, to compare relative stability between two phases, one can use their cohesive energies if they have their corresponding single configurations. Since both \textit{T} and \textit{H} phases of any TMDCs are determined by their corresponding single equilibrium configurations, their cohesive energies may be good indicators to determine the relative stability.

Another interesting common trend was observed in the structural
stiffness. As shown in Fig.~\ref{Fig2}, the energy curves for all the
\textit{T} phases are flatter than their \textit{H} counterparts
entailing that the former phases are softer than the latter. Such
stiffness behavior can be observed more profoundly in tensile strain
region, where $a$ is larger than the equilibrium lattice constant
$a_\mathrm{eq}$. Because of this stiffness trend, external stress
applied on M$_4^P$X$_2$ (M in group 4) may not allow by itself their
phase transition from their equilibrium \textit{T} phase to their
\textit{H} counterpart. For M$_5^P$X$_2$ (M in group 5), which can be
in either phase near equilibrium, we expect the compounds in the
\textit{H} phase may be converted into the \textit{T} phase by
external tensile stress, but not the other way around. For
M$_6^P$X$_2$ (M in group 6), the energy curves representing two phases
cross each other not only at a certain tensile strain value, but also
at a certain compressive strain value, implying that external tensile
(or compressive) stress may induce nearly spontaneous phase transition
from their equilibrium \textit{H} phase to their \textit{T}
counterparts, as we further discuss later. Figure~\ref{Fig2} also
shows that the equilibrium lattice constant $a_{\textrm{eq}}$ of each
\textit{T} phase tends to be almost the same as or a little larger
than that of its \textit{H} counterpart, with the exception of
CrX$_2$, whose \textit{T}-phase $a_{\textrm{eq}}$ is much larger than
that of its \textit{H} phase. 

\begin{figure*}[t]
\includegraphics[width=1.0\textwidth]{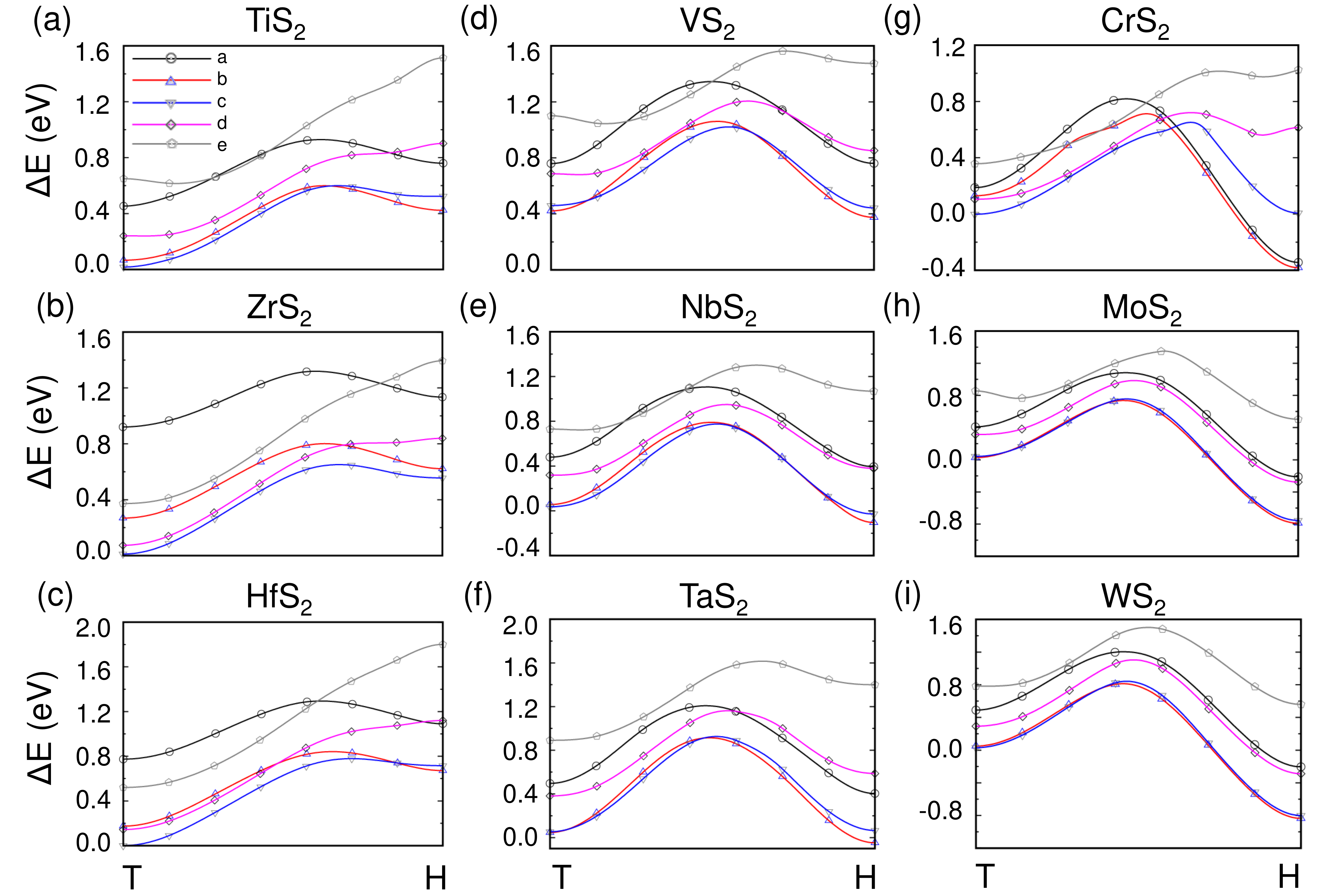}
\caption{(Color online)
Relative change in energy during the transition between the
\textit{T} and the \textit{H} phases of all the sulfides: (a) TiS$_2$ 
(b) ZrS$_2$, (c) HfS$_2$, (d) VS$_2$, (e) NbS$_2$, (f) TaS$_2$, (g)
CrS$_2$, (h) MoS$_2$ and (i) WS$_2$ with various lattice constants
indicated by $^{a)}$ through $^{e)}$, which correspond to in-plane
strain values with the same indices listed in Tables~\ref{T1} through
\ref{T3} for sulfides. The relative energy values in each graph are
referenced to the energy of the equilibrium $T$ phase of given MS$_2$.
The columns and rows of the graphs are arranged according to the
groups and periods of the transition metal elements. 
\label{Fig3}}
\end{figure*}

\begin{figure*}[t]
\includegraphics[width=1.0\textwidth]{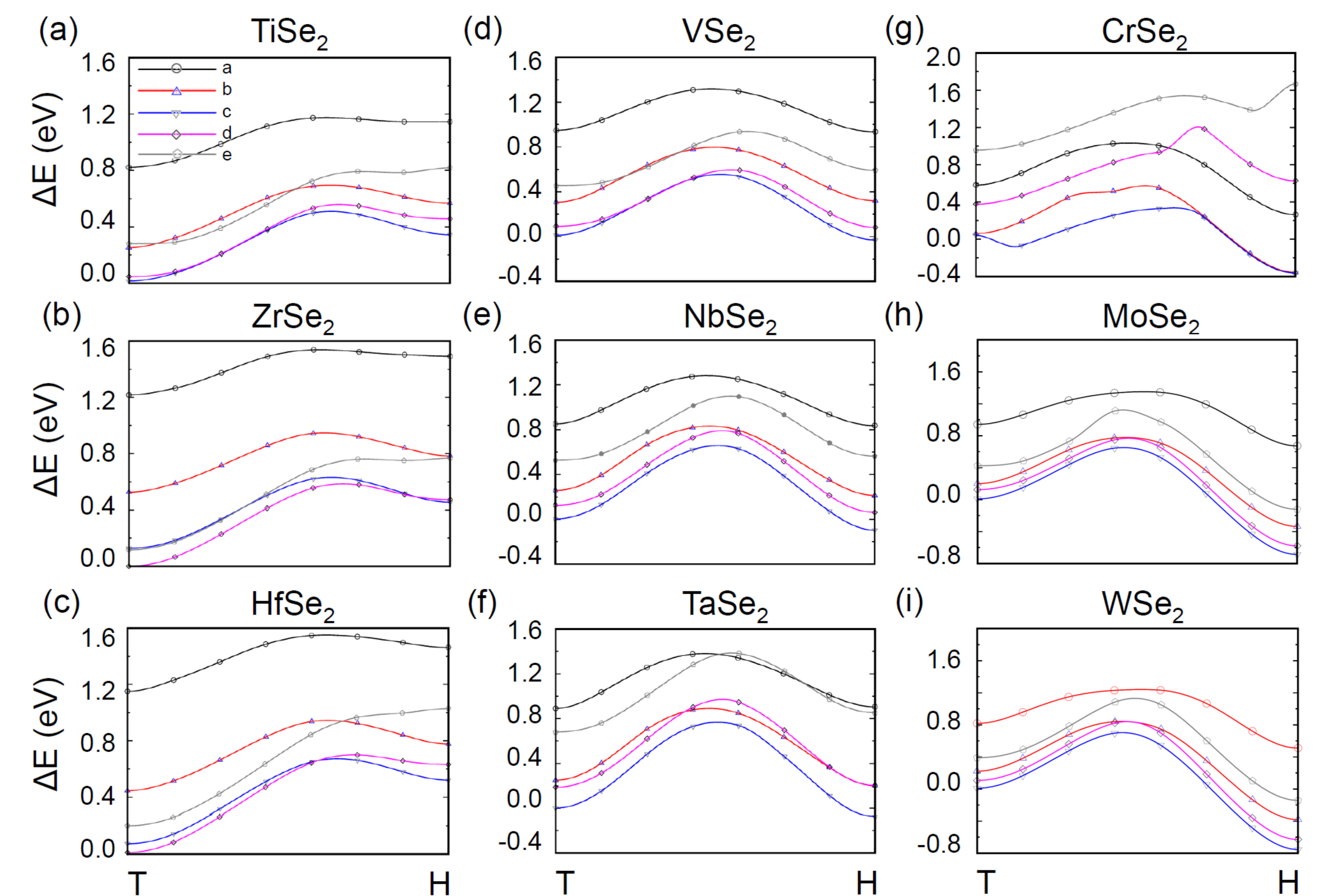}
\caption{(Color online)
Relative change in energy during the transition between the
\textit{T} and the \textit{H} phases of all the selenides: (a)
TiSe$_2$ (b) ZrSe$_2$, (c) HfSe$_2$, (d) VSe$_2$, (e) NbSe$_2$, (f)
TaSe$_2$, (g) CrSe$_2$, (h) MoSe$_2$ and (i) WSe$_2$ with various
lattice constants indicated by $^{a)}$ through $^{e)}$, which
correspond to in-plane strain values with the same indices listed in
Tables~\ref{T1} through \ref{T3} for selenides. The relative energy
values in each graph are referenced to the energy of the equilbrium
$T$ phase of given MSe$_2$. The columns and rows of the graphs are
arranged according to the groups and periods of the transition metal
elements.
\label{Fig4}}
\end{figure*}

\begin{landscape}
\begin{table*}[bt]
\caption{
For every M$_4^P$X$_2$ (X$=$S, Se) with several different lattice
constants ($a$), the corresponding in-plane strain values
($\epsilon$), activation barriers (${\Delta}E_{\textrm{b}}$) and
reaction energies (${\Delta}E_{\textrm{r}}$) are summarized. The same
labels $^{a)}$ through $^{e)}$ were used as used in Figs.~\ref{Fig3}
and \ref{Fig4}. All the energy values are referenced to the energy of
their respective $T$ phases. ``$-$'' marked in
${\Delta}E_{\textrm{b}}$ indicates that the energy increases
monotonically during the corresponding phase transition without the
energy maximum.
\label{T1}}
\begin{tabular}{c|c*{3}{|c*{3}{D{.}{.}{-1}}}}
\hline\hline
\multicolumn{2}{c|}{M} & \multicolumn{4}{c|}{Ti} & \multicolumn{4}{c|}{Zr} & \multicolumn{4}{c}{Hf} \\
\hline
X & $a$~(\AA) & \multicolumn{2}{c}{$\epsilon$~(\%)} & \multicolumn{1}{c}{${\Delta}E_{\rm{b}}$~(eV)} & \multicolumn{1}{c|}{${\Delta}E_{\rm{r}}$~(eV)} & \multicolumn{2}{c}{$\epsilon$~(\%)} & \multicolumn{1}{c}{${\Delta}E_{\rm{b}}$~(eV)} & \multicolumn{1}{c|}{${\Delta}E_{\rm{r}}$~(eV)} & \multicolumn{2}{c}{$\epsilon$~(\%)} & \multicolumn{1}{c}{${\Delta}E_{\rm{b}}$~(eV)} & \multicolumn{1}{c}{${\Delta}E_{\rm{r}}$~(eV)} \\ 
\hline
  \multirow{6}{*}{S} & 3.12 & $^{a)}$& -8.34 & 0.48 & 0.31 &     &           &      &      &     &            &      &    \\
                     & 3.29 & $^{b)}$& -3.24 & 0.53 & 0.36 & $^{a)}$&  10.72 & 0.40 & 0.21 & $^{a)}$&  -9.52 & 0.52 & 0.32\\
                     & 3.46 & $^{c)}$&  1.85 & 0.58 & 0.51 & $^{b)}$&  -0.60 & 0.53 & 0.35 & $^{b)}$&  -4.75  & 0.66 & 0.50\\
                     & 3.64 & $^{d)}$&  6.94 & \multicolumn{1}{c}{$-$} & 0.66 & $^{c)}$&  -1.32 & 0.64 & 0.55 & $^{c)}$&  0.01 & 0.78 & 0.72\\
                     & 3.81 & $^{e)}$& 12.03 & \multicolumn{1}{c}{$-$} & 0.87 & $^{d)}$&  3.38 & \multicolumn{1}{c}{$-$} & 0.77 & $^{d)}$&  4.77 & \multicolumn{1}{c}{$-$} & 0.98\\
                     & 3.98 &      &         &      &      & $^{e)}$&  8.08 & \multicolumn{1}{c}{$-$} & 1.02 & $^{e)}$&  9.53 & \multicolumn{1}{c}{$-$} & 1.28\\
\hline
  \multirow{6}{*}{Se} & 3.12 &$^{a)}$& -11.58 & 0.35 & 0.32 &       &         &      &      &       &        &      &     \\
                      & 3.29 &$^{b)}$&  -6.67 & 0.44 & 0.32 & $^{a)}$& -13.37 & 0.32 & 0.28 &$^{a)}$& -12.46 & 0.40 & 0.31\\
                      & 3.46 &$^{c)}$&  -1.76 & 0.49 & 0.33 & $^{b)}$&  -8.81 & 0.42 & 0.26 &$^{b)}$&  -7.85 & 0.50 & 0.33\\
                      & 3.64 &$^{d)}$&  3.16 & 0.51 & 0.41 & $^{c)}$&  -4.24 & 0.50 & 0.33 &$^{c)}$&  -3.25 & 0.60 & 0.45\\
                      & 3.81 &$^{e)}$&  8.07 & 0.51 & 0.54 & $^{d)}$&  0.31 & 0.59 & 0.47 &$^{d)}$&  1.36 & 0.69 & 0.62\\
                      & 3.98 &      &         &      &      & $^{e)}$&   4.16 & 0.64 & 0.65 &$^{e)}$&   5.97 & \multicolumn{1}{c}{$-$} & 0.82\\
\hline\hline
\end{tabular}
\end{table*}
\end{landscape}

\begin{landscape}
\begin{table*}[bt]
\caption{
For every M$_5^P$X$_2$ (X$=$S, Se) with several different lattice
constants ($a$), the corresponding in-plane strain values
($\epsilon$), activation barriers (${\Delta}E_{\textrm{b}}$) and
reaction energies (${\Delta}E_{\textrm{r}}$) are summarized. The same
labels $^{a)}$ through $^{e)}$ were used as denoted in Figs.~\ref{Fig3}
and \ref{Fig4}. All the reaction energy values are referenced to the
energy of their respective $H$ phases. Negative activation barrier
values indicate that the corresponding \textit{T} phase is more
stable than its reference \textit{H} phase.
\label{T2}}
\begin{tabular}{c|c*{3}{|c*{3}{D{.}{.}{-2}}}}
\hline\hline
\multicolumn{2}{c|}{M} & \multicolumn{4}{c|}{V} & \multicolumn{4}{c|}{Nb} & \multicolumn{4}{c}{Ta} \\
\hline
X & $a$~(\AA) & \multicolumn{2}{c}{$\epsilon$~(\%)} & \multicolumn{1}{c}{${\Delta}E_{\rm{b}}$~(eV)} & \multicolumn{1}{c|}{${\Delta}E_{\rm{r}}$~(eV)} & \multicolumn{2}{c}{$\epsilon$~(\%)} & \multicolumn{1}{c}{${\Delta}E_{\rm{b}}$~(eV)} & \multicolumn{1}{c|}{${\Delta}E_{\rm{r}}$~(eV)} & \multicolumn{2}{c}{$\epsilon$~(\%)} & \multicolumn{1}{c}{${\Delta}E_{\rm{b}}$~(eV)} & \multicolumn{1}{c}{${\Delta}E_{\rm{r}}$~(eV)} \\ 
\hline
  \multirow{6}{*}{S} & 2.94 &$^{a)}$& -7.41 & 0.58 & 0.00 &              &  &    &      &   &     &      &      \\
                     & 3.12 &$^{b)}$&  -1.96 & 0.69 & 0.05 &$^{a)}$&  -7.23 & 0.71 & 0.09 &$^{a)}$& -6.55 & 0.80 & 0.09 \\
                     & 3.29 &$^{c)}$&  3.48 & 0.58 & 0.02 &$^{b)}$&  -2.08 & 0.90 & 0.16 &$^{b)}$& -1.36 & 0.96 & 0.09 \\
                     & 3.46 &$^{d)}$&   8.93 & 0.35 & -0.17 &$^{c)}$&  3.07 & 0.80 & 0.06 &$^{c)}$&  3.83 & 0.86 & -0.01 \\
                     & 3.64 &$^{e)}$&  14.38 & 0.09 & -0.37 &$^{d)}$&   8.22 & 0.57 & -0.06 &$^{d)}$&  9.02 & 0.57 & -0.21 \\
                     & 3.81 &    &         &      &      &$^{e)}$& 13.38 & 0.23 & -0.34 &$^{e)}$& 14.21 & 0.21 & -0.51 \\
\hline
  \multirow{6}{*}{Se} & 2.94 &$^{a)}$&  -11.57 & 0.38 & 0.01 &  &    &    &      &      &              &      &       \\
                      & 3.12 &$^{b)}$&  -6.36 & 0.48 & -0.01 &$^{a)}$& -10.23 & 0.44 & 0.01 &$^{a)}$&  -6.55 & 0.48 & -0.01 \\
                      & 3.29 &$^{c)}$&  -1.16 & 0.58 & 0.04 &$^{b)}$&  -5.66 & 0.62 & 0.04 &$^{b)}$&  -1.36 & 0.69 & 0.05 \\
                      & 3.46 &$^{d)}$&   4.04 & 0.51 & 0.01 &$^{c)}$&  -0.69 & 0.76 & 0.10 &$^{c)}$&  3.83 & 0.84 & 0.08 \\
                      & 3.64 &$^{e)}$&  9.24 & 0.34 & -0.14 &$^{d)}$&  4.27 & 0.73 & 0.06 &$^{d)}$&   9.02 & 0.76 & -0.02 \\
                      & 3.81 &       &         &      &       &$^{e)}$&   9.24 & 0.58 & -0.04 &$^{e)}$&  14.21 & 0.53 & -0.17 \\
\hline\hline
\end{tabular}
\end{table*}
\end{landscape}

\begin{landscape}
\begin{table*}[bt]
\caption{
For every M$_6^P$X$_2$ (X$=$S, Se) with several different lattice
constants ($a$), the corresponding in-plane strain values
($\epsilon$), activation barriers (${\Delta}E_{\textrm{b}}$) and
reaction energies (${\Delta}E_{\textrm{r}}$) are summarized. The same
labels $^{a)}$ through $^{e)}$ were used as used in Figs.~\ref{Fig3}
and \ref{Fig4}. The symbols $^\dagger$ and $^\ddagger$ correspond to
the same symbols used in Fig.~\ref{Fig5}. All the energy values are
referenced to the energy of their respective $H$ phases. Negative
activation barrier values indicate that the corresponding \textit{T}
phase is more stable than its reference \textit{H} phase.
\label{T3}}
\begin{tabular}{c|c*{3}{|c*{3}{D{.}{.}{-2}}}}
\hline\hline
\multicolumn{2}{c|}{M} & \multicolumn{4}{c|}{Cr} & \multicolumn{4}{c|}{Mo} & \multicolumn{4}{c}{W} \\
\hline
X & $a$~(\AA) & \multicolumn{2}{c}{$\epsilon$~(\%)} & \multicolumn{1}{c}{${\Delta}E_{\rm{b}}$~(eV)} & \multicolumn{1}{c|}{${\Delta}E_{\rm{r}}$~(eV)} & \multicolumn{2}{c}{$\epsilon$~(\%)} & \multicolumn{1}{c}{${\Delta}E_{\rm{b}}$~(eV)} & \multicolumn{1}{c|}{${\Delta}E_{\rm{r}}$~(eV)} & \multicolumn{2}{c}{$\epsilon$~(\%)} & \multicolumn{1}{c}{${\Delta}E_{\rm{b}}$~(eV)} & \multicolumn{1}{c}{${\Delta}E_{\rm{r}}$~(eV)} \\ 
\hline
  \multirow{5}{*}{S} & 2.94 &$^{a)}$& -3.03 & 1.16 & 0.53 &$^{a)}$& -7.68 & 1.29 & 0.62  &$^{a)}$& -7.74 & 1.41 & 0.69 \\
                     & 3.12 &$^{b)}$& 2.61 & 0.60 & -0.51 &$^{b)}$& -2.25 & 1.53 & 0.82  &$^{b)}$& -2.31 & 1.65 & 0.89 \\
                     & 3.29 &$^{c)}$& 8.38 & 0.65 & 0.01 &$^{c)}$& 3.18 & 1.51 & 0.80 &$^{c)}$& 3.11 & 1.64 & 0.83 \\
                     & 3.46 &$^{d)}$& 14.09 & 0.62 & 0.51 &$^{d)}$& 8.61 & 1.26 & 0.59 &$^{d)}$& 8.54 & 1.39 & 0.58 \\
                     & 3.64 &$^{e)}$& 19.79 & 0.04 & -0.67 &$^{e)}$& 14.04 & 0.85 & 0.35  &$^{e)}$& 13.97 & 0.94 & 0.22 \\
                     & 3.81 &      &       &      &   &$^\dagger$& 19.47 & 0.31 & -0.34 &$^\dagger$ & 19.29 & 0.43 & -0.29 \\
                     & 3.98 &      &       &      &   &$^\ddagger$ & 24.80 & 0.01 & -0.97 &$^\ddagger$ & 24.62 & 0.06 & -1.04 \\
\hline
  \multirow{5}{*}{Se} & 2.94 &$^{a)}$& -8.25 & 0.77 & 0.32 &$^{a)}$& -11.48 & 0.68 & 0.27 &$^{a)}$& -11.47 & 0.73 & 0.31 \\
                     & 3.12 &$^{b)}$& -2.85 & 0.93 & 0.42 &$^{b)}$& -6.27 & 1.11 & 0.54 &$^{b)}$& -6.26 & 1.22 & 0.60 \\
                     & 3.29 &$^{c)}$& 2.55 & 0.70 & 0.41 &$^{c)}$& -1.06 & 1.34 & 0.70 &$^{c)}$& -1.05 & 1.46 & 0.76 \\
                     & 3.46 &$^{d)}$& 7.94 & 0.58 & -0.25 &$^{d)}$& 4.15 & 1.35 & 0.70 &$^{d)}$& 4.16 & 1.47 & 0.74 \\
                     & 3.64 &$^{e)}$& 13.34 & 0.16 & -0.43 &$^{e)}$& 9.35 & 1.24 & 0.55 &$^{e)}$& 9.37 & 1.27 & 0.53 \\
                     & 3.81 &    &          &      &   &$^\dagger$ & 14.46 & 0.74 & 0.26  &$^\dagger$ & 14.48 & 0.95 & 0.23 \\ 
                     & 3.98 &    &          &      &   &$^\ddagger$ & 19.56 & 0.29 & -0.32 &$^\ddagger$ & 19.59 & 0.40 & -0.28 \\
\hline\hline
\end{tabular}
\end{table*}
\end{landscape}

To study the strain effect on the phase transitions between \textit{T}
and \textit{H} phases of TMDCs, we simulated the phase transition
process of TMDC materials by sliding one of two chalcogen planes, 
which sandwich the transition metal plane, 
while applying compressive and tensile strain or adjusting lattice constant $a$ keeping the hexagonal lattice. (See Fig.~S1(a) in Supplementary Information.) 
For each transition, we evaluated the relative energy
with respect to the corresponding \textit{T} phase equilibrium
configuration as a function of the relative coordinates of the shifted
chalcogen atom. Our results for all M$_G^P$X$_2$ materials with X$=$S
and Se are shown in Figs.~\ref{Fig3} and \ref{Fig4}, respectively,
where each graph represents several relative energy curves plotted for
different biaxial strains obtained by adjusting the lattice constant.

From these results, we extracted the corresponding in-plane strain
values ($\epsilon$), activation barriers (${\Delta}E_{\textrm{b}}$)
and reaction energies (${\Delta}E_{\textrm{r}}$) are summarized in
Tables~\ref{T1} through \ref{T3}. For each given lattice constant or
in-plane strain in each M$_G^P$X$_2$, ${\Delta}E_{\textrm{b}}$ values
were evaluated by the energy difference from the lowest energy phase,
either \textit{T} or \textit{H} phase, to the highest energy state
between two phases, while ${\Delta}E_{\textrm{r}}$ were obtained by
the energy differences between two phases.

As clearly seen in Figs.~\ref{Fig3} and \ref{Fig4}, and from
Tables~\ref{T1} through \ref{T3}, $E_\textrm{b}$ values depend
sensitively on change in the lattice constant $a$ or on the
corresponding in-layer strain $\epsilon$. According to our
calculation, M$_4^P$X$_2$ (X=S, Se) materials (M in group 4) in the
$T$ phase become more stable than those in the $H$ phase as shown in
Fig.~\ref{Fig3} (a--c) and Fig.~\ref{Fig4} (a--c). Moreover, we found
that the larger tensile strain, the higher activation energy barrier
and reaction energy during the transition to the \textit{H} phase as
listed in Table~\ref{T1}, because of difference in their structural
stiffness as already mentioned above. This indicates that any tensile
stress may keep any TMDCs with a metal atom belonging to the $G=4$
group in the \textit{T} phase, but they may experience the phase
transition to the \textit{H} phase at large compressive strain values.

Different transition behaviors were observed for M$_G^P$X$_2$ (X=S,
Se) with $G=5$ and 6. It was found that for each case of $G=5$,
either one of two phases is almost as stable as the other near the
equilibrium and up to moderate compressive strain values, resulting in
small reaction energies and admissible activation energy barriers,
which depend less sensitively on strain values, as shown in
Figs.~\ref{Fig3} (d--f) and \ref{Fig4} (d--f), and as summarized in
Table~\ref{T2}, than for the other cases of $G=4$ and $G=6$ shown in
Tables~\ref{T1} and \ref{T3}. 
Our result is in an agreement with a previous study on VS$_2$ showing that small strain induces a phase transition from its \textit{H} phase to its \textit{T} counterpart.~\cite{Kan2014}

The most intriguing transition behaviors under external stress were
found in M$_6^P$X$_2$ (M in group 6). Each material in this category
is in the \textit{H} phase near equilibrium, but as external stress
increases, its \textit{T} phase becomes more stable than its
\textit{H} counterpart. Such phase crossover is clearly seen in
Fig.~\ref{Fig2} (c) and (f), especially in the region of
$a>a_\mathrm{eq}$ (tensile strain) for all cases and even in the
region of $a<a_\mathrm{eq}$ (compressive strain) for a case of
WSe$_2$.

\begin{figure}[t]
\includegraphics[width=1.0\columnwidth]{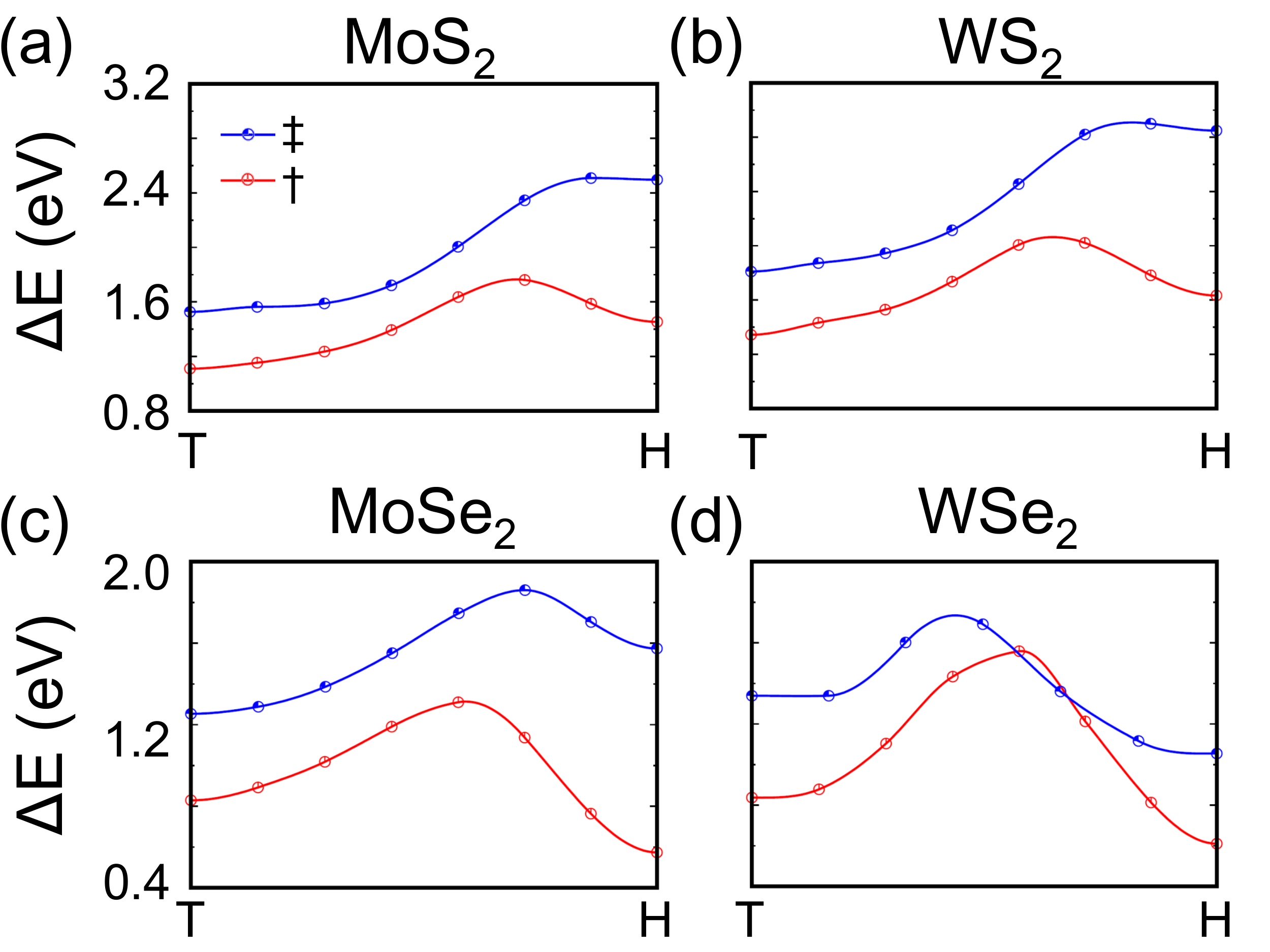}
\caption{(Color online)
Relative change in energy during the transition between the \textit{T}
and the \textit{H} phases of (a) MoS$_2$, (b) WS$_2$ (c) MoSe$_2$, and
(d) WSe$_2$ at two large lattice constants, $a=3.81$~\AA (red) and
3.98~\AA (blue) or large tensile strain values, indicated by
$^\dagger$ and $^\ddagger$, which correspond to in-plane strain values
with the same symbols listed in Table~\ref{T3}.
\label{Fig5}}
\end{figure}

As shown in Figs.~\ref{Fig3} (g--i), \ref{Fig4} (g--i), and
\ref{Fig5}, and listed in Table~\ref{T3}, the activation energy
barrier becomes smaller with tensile strain applied although strain
requires extra energy cost. Such energy barrier becomes even nearly
zero at 20~\% or more in tensile strain, resulting in nearly
spontaneous phase transition. Thanks to the bonding flexibility of S 
or Se atoms in TMDCs, the large tensile strain up to 20~\% may be
achieved,~\cite{Li2012} as similarly observed in Mo-S-I
nanowires.~\cite{{Kang2010}, {Kang2010a}} Based on energy curves shown
in Figs.~\ref{Fig2} (c, f), and \ref{Fig5}, we further estimated the
energy cost for ${\sim}20$~\% strain to be about 0.33 eV per each Mo-S
bond, which corresponds to ${\sim}33$~GPa. 
We also noted, however, the structural stability under large tensile 
strain applied may not be guaranteed with a small unit cell as used 
in this study. Thus, we prepared much larger $6\times6$ supercell 
for MoS$_2$ to realistically check the stability under large tensile 
strain. It was found that MoS$_2$ maintains its symmetry and structure
under tensile strain as large as near 20~\%.
This may provide a feasible microscopic explanation on experimental observations of pressure-induced insulator-metal transition of MoS$_2$ implying the phase transition from its semiconducting \textit{H} phase to its metallic \textit{T} phase induced by strain.~\cite{{Suenaga2014},{Lin2014}}

To verify that our results on structural stability and strain dependence of phase transition are reliable, we further calculated with LDA. The relative energies calculated with LDA fall within only a few to ten percent ranges of those with GGA. In addition, we also considered SOC effect for heavy transition metal atoms, such as Hf, Ta, and W. For WS$_2$, for instance, the equilibrium lattice constants of both H and T phases are essentially the same regardless of whether calculated with or without SOC. Moreover, the SOC effect does not change our finding that its H phase is more stable than its T counterpart, although the energy difference between two phases decreases significantly with SOC resulting in a narrow range of lattice constants where the H phase is more stable than the T phase. (See Fig.~S1 (b) in Supplementary Information.) This means that phase transition could be achievable with smaller strain values than those we estimated above.

\begin{figure*}[t]
\includegraphics[width=1.0\textwidth]{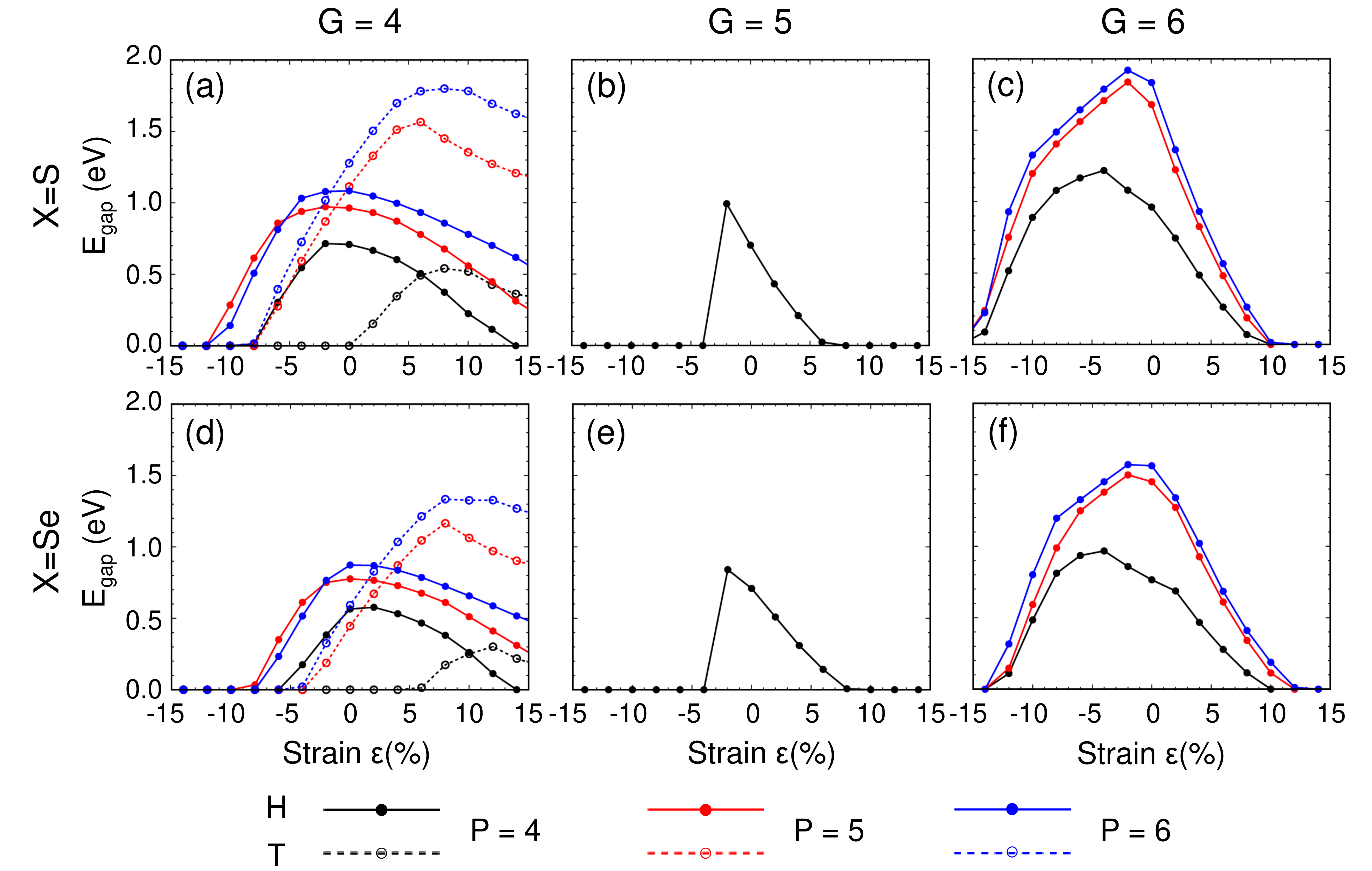}
\caption{(Color online) Electronic band gap as function of strain in
(a) M$_4^P$S$_2$ (b) M$_5^P$S$_2$, (c) M$_6^P$S$_2$, (d)
M$_4^P$Se$_2$, (e) M$_5^P$Se$_2$, and (f) M$_6^P$Se$_2$. Black, red,
and blue lines in each graph represents its band gap changes for
$P=4$, 5, and 6, respectively, that is, Ti, Zr, and Hf for $G=4$; V,
Nb, and Ta for $G=5$; and Cr, Mo, and W for $G=6$. Solid lines
indicate the band gap of the \textit{H} phases and dotted lines is
\textit{T}.
\label{Fig6}}
\end{figure*}

We further investigate the strain effects on the electronic properties
of all MX$_2$ materials in the \textit{H}- and the \textit{T}-phases.
Earlier studies showed that the \textit{H} phase of MoS$_2$ is
semiconducting,~\cite{Mattheiss1973} while its \textit{T} counterpart
is metallic.~\cite{Schollhorn1992} Figure~\ref{Fig6} presents our
results for the strain dependence of the electronic band gap of
various MX$_2$ materials. We divided our results into sulfides 
(a--c) and selenides (d--f), and collected them group by group 
to which the metal atoms belong, i.e. M in the group 4 
($G=4$, Ti, Zr, Hf) in (a) and (d), the group 5 ($G=5$, V, Nb, Ta) in
(b) and (e), and the group 6 ($G=6$, Cr, Mo, W) in (c) and (f). 

Our calculation revealed that as shown in Fig.~\ref{Fig6} (a, d) for
the group 4, not only the \textit{H} phase, but also the equilibrium
\textit{T} phase are semiconducting near their equilibria except for
TiS$_2$ and TiSe$_2$, whose equilibrium \textit{T} phases are
metallic. Intriguingly, even very small tensile strain triggers the
metal-semiconductor transition in the \textit{T}-phase TiS$_2$,
whereas it remains metallic under compressive strain. The
\textit{T}-phase TiSe$_2$ becomes semiconducting at tensile strain
above ${\sim}5$~\%. In the other two metal cases (M=Zr and Hf) in the
group 4, their \textit{T} phases exhibit even larger energy band gap
than their \textit{H} counterparts, regardless of whether X=S or Se.
Overall, the maximum band gap values of M$_4^P$X$_2$ fall within the
ranges of $0.5-1.1$~eV for the \textit{H} phases near equilibrium or
under small compressive strain, and of $1.2-1.8$~eV for the \textit{T}
phases near ${\sim}8$~\% of tensile strain values. All the materials in
this category eventually experience semiconducting-metal transitions
under compressive strains. 

For the group 5 ($G=5$), all the M$_5^P$X$_2$ materials remain
metallic under any strain values, whatever the chalcogen element X is,
except for VS$_2$ and VSe$_2$ in the \textit{H} phase. Both exceptions
are semiconducting within the strain range of
$-5$~\%$<\epsilon\lesssim+8$~\% with the maximum energy band gap of 
around 1~eV.

The group 6 ($G=6$) cases exhibit interesting phase-dependent
electronic characteristics: all the \textit{T}-phase M$_6^P$X$_2$
materials remain metallic at any applied strain, while their $H$-phase
counterparts experience their gap opening and closing as the applied
strain increases from $-15$~\% to $+15$~\%, as shown in
Fig.~\ref{Fig6} (c) and (f). They have the maximum band gaps near
$\epsilon=0$~\%.
To verify whether our calculated electronic structures are reliable, we performed more accurate calculation on MoS$_2$ based on self-energy correction within the \textit{GW} approximation. We found that even in \textit{GW} calculations, its \text{T} phase remains metallic and its \textit{H} phase does semiconducting at their corresponding equilibria. For the latter case, its DFT-based band gap was underestimated by ${\sim}1.4$~eV from its \textit{GW} value. (See Fig.~S2~(a) in Supplementary Information.) Nonetheless, it turned out that band gap trends are similar between DFT and \textit{GW} calculations. Like the GGA result shown in Fig.~\ref{Fig6} (c), the \textit{GW} band gap begins to close at ${\sim}14$~\% of tensile strain, which is larger than ${\sim}10$~\% in GGA. (See Fig.~S2~(b) in Supplementary Information.) This indicates our observed insulator-metal transition may be robust and be induced by tensile strain.

\section{Conclusions}
\label{Summary}
In summary, we investigated the structural and electronic properties
of MX$_2$ (M=Ti, V, Cr, Zr, Nb, Mo, Hf, Ta, or W; X=S or Se) and its
phase transition using density functional theory (DFT). We found that
each group to which M belongs in the periodic table determines its
characteristic equilibrium phase. The stable structural phases of
group 4 (Ti, Zr, Hf) are mainly octahedral, while those of group 6
(Cr, Mo, W) are trigonal. Interestingly, those of group 5 (V, Nb, Ta)
can be either one, since their two phase structures have almost the
same total energy near their corresponding equilibrium lattice
constants.

We also explored phase transitions between octahedral-\textit{T} and
trigonal-\textit{H} phases for all MX$_2$ compositions by evaluating
the relative stability during their transition processes under various
strain conditions. This exploration yielded the activation energy
barriers between two phases as well as the reaction energies. Our
results revealed that the compounds with M in the group 4 would
experience phase transitions from their equilibrium \textit{T} phases
to their \textit{H} counterpart under mainly compressive strains for
X=S and any strain conditions for X=Se. However, such transitioned
phases would be converted back to their original equailibrium phases.
Compounds with M in the group 5, which can exist in either the
\textit{T} or the \textit{H} phase near their equilibrium or under
compressive stress and would experience the phase transitions between
two phases, tend to form the \textit{H} phases more easily under
larger tensile stresses. Most interesting structural phase transitions
were observed in compounds with M in the group 6, which form the
\textit{H} phases at their equilibria and hardly experience their
phase transition to the \textit{T} phases due to large energy barrier
values, but large tensile strain may make their transition to the
\textit{T} phases relatively easily, and for some cases such
transitions occur almost spontaneously with very small energy
barriers. It was, moreover, found that external stress can be used to
control the electronic properties, such as electronic energy gap, and
eventually metal-semiconductor transition. Based on these results, we
can surmise that the electron-phonon coupling may be an important
property in some of MX$_2$ materials, which will be considered as our
future work.

\section*{Acknowledgments}
We acknowledge financial support from the Korean government through
National Research Foundation (NRF-2015R1A2A2A01006204, 2019R1A2C1005417)
and the Ministry of Trade, Industry \& Energy (MOTIE) of Korea (Project
No.  10045360). Some portion of our computational work was done using
the resources of the KISTI Supercomputing Center (KSC-2016-C3-0034).


\end{document}